\documentclass[twocolumn,aps,showpacs,superscriptaddress]{revtex4}
\usepackage{graphicx}% Include figure files
\usepackage{dcolumn}% Align table columns on decimal point
\usepackage{bm}% bold math
\begin{document}
% A useful Journal macro
\def\Journal#1#2#3#4{{#1} {\bf #2}, #3 (#4)}

% Some useful journal names
\def\NCA{Nuovo Cimento}
\def\NIM{Nucl. Instr. Meth.}
\def\NIMA{{Nucl. Instr. Meth.} A}
\def\NPB{{Nucl. Phys.} B}
\def\NPA{{Nucl. Phys.} A}
\def\PLB{{Phys. Lett.}  B}
\def\PRL{Phys. Rev. Lett.}
\def\PRC{{Phys. Rev.} C}
\def\PRD{{Phys. Rev.} D}
\def\ZPC{{Z. Phys.} C}
\def\JPG{{J. Phys.} G}
\def\CPC{Comput. Phys. Commun.}
\def\EPJ{{Eur. Phys. J.} C}

\preprint{}
\title{Pion, kaon, proton and anti-proton transverse momentum distributions from p+p and d+Au collisions at $\sqrt{s_{_{NN}}} = 200$ GeV}
\affiliation{Argonne National Laboratory, Argonne, Illinois 60439}
\affiliation{University of Bern, 3012 Bern, Switzerland}
\affiliation{University of Birmingham, Birmingham, United Kingdom}
\affiliation{Brookhaven National Laboratory, Upton, New York
11973} \affiliation{California Institute of Technology, Pasadena,
California 91125} \affiliation{University of California, Berkeley,
California 94720} \affiliation{University of California, Davis,
California 95616} \affiliation{University of California, Los
Angeles, California 90095} \affiliation{Carnegie Mellon
University, Pittsburgh, Pennsylvania 15213} \affiliation{Creighton
University, Omaha, Nebraska 68178} \affiliation{Nuclear Physics
Institute AS CR, 250 68 \v{R}e\v{z}/Prague, Czech Republic}
\affiliation{Laboratory for High Energy (JINR), Dubna, Russia}
\affiliation{Particle Physics Laboratory (JINR), Dubna, Russia}
\affiliation{University of Frankfurt, Frankfurt, Germany}
\affiliation{Institute  of Physics, Bhubaneswar 751005, India}
\affiliation{Indian Institute of Technology, Mumbai, India}
\affiliation{Indiana University, Bloomington, Indiana 47408}
\affiliation{Institut de Recherches Subatomiques, Strasbourg,
France} \affiliation{University of Jammu, Jammu 180001, India}
\affiliation{Kent State University, Kent, Ohio 44242}
\affiliation{Lawrence Berkeley National Laboratory, Berkeley,
California 94720} \affiliation{Massachusetts Institute of
Technology, Cambridge, MA 02139-4307}
\affiliation{Max-Planck-Institut f\"ur Physik, Munich, Germany}
\affiliation{Michigan State University, East Lansing, Michigan
48824} \affiliation{Moscow Engineering Physics Institute, Moscow
Russia} \affiliation{City College of New York, New York City, New
York 10031} \affiliation{NIKHEF, Amsterdam, The Netherlands}
\affiliation{Ohio State University, Columbus, Ohio 43210}
\affiliation{Panjab University, Chandigarh 160014, India}
\affiliation{Pennsylvania State University, University Park,
Pennsylvania 16802} \affiliation{Institute of High Energy Physics,
Protvino, Russia} \affiliation{Purdue University, West Lafayette,
Indiana 47907} \affiliation{University of Rajasthan, Jaipur
302004, India} \affiliation{Rice University, Houston, Texas 77251}
\affiliation{Universidade de Sao Paulo, Sao Paulo, Brazil}
\affiliation{University of Science \& Technology of China, Anhui
230027, China} \affiliation{Shanghai Institute of Applied Physics,
Shanghai 201800, China} \affiliation{SUBATECH, Nantes, France}
\affiliation{Texas A\&M University, College Station, Texas 77843}
\affiliation{University of Texas, Austin, Texas 78712}
\affiliation{Tsinghua University, Beijing 100084, China}
\affiliation{Valparaiso University, Valparaiso, Indiana 46383}
\affiliation{Variable Energy Cyclotron Centre, Kolkata 700064,
India} \affiliation{Warsaw University of Technology, Warsaw,
Poland} \affiliation{University of Washington, Seattle, Washington
98195} \affiliation{Wayne State University, Detroit, Michigan
48201} \affiliation{Institute of Particle Physics, CCNU (HZNU),
Wuhan 430079, China} \affiliation{Yale University, New Haven,
Connecticut 06520} \affiliation{University of Zagreb, Zagreb,
HR-10002, Croatia}

\author{J.~Adams}\affiliation{University of Birmingham, Birmingham, United Kingdom}
\author{M.M.~Aggarwal}\affiliation{Panjab University, Chandigarh 160014, India}
\author{Z.~Ahammed}\affiliation{Variable Energy Cyclotron Centre, Kolkata 700064, India}
\author{J.~Amonett}\affiliation{Kent State University, Kent, Ohio 44242}
\author{B.D.~Anderson}\affiliation{Kent State University, Kent, Ohio 44242}
\author{D.~Arkhipkin}\affiliation{Particle Physics Laboratory (JINR), Dubna, Russia}
\author{G.S.~Averichev}\affiliation{Laboratory for High Energy (JINR), Dubna, Russia}
\author{S.K.~Badyal}\affiliation{University of Jammu, Jammu 180001, India}
\author{Y.~Bai}\affiliation{NIKHEF, Amsterdam, The Netherlands}
\author{J.~Balewski}\affiliation{Indiana University, Bloomington, Indiana 47408}
% ...
%\author{
%\begin{center}(STAR Collaboration)\end{center}
%}
\collaboration{STAR Collaboration}\homepage{www.star.bnl.gov}\noaffiliation

%\affiliation{TOF}

\date{\today}% It is always \today, today,
\begin{abstract}
Identified mid-rapidity particle spectra of $\pi^{\pm}$,
$K^{\pm}$, and $p(\bar{p})$ from 200 GeV p+p and d+Au collisions
are reported. A time-of-flight detector based on multi-gap
resistive plate chamber technology is used for particle
identification. The particle-species dependence of the Cronin
effect is observed to be significantly smaller than that at lower
energies. The ratio of the nuclear modification factor ($R_{dAu}$)
between protons ($p+\bar{p}$) and charged hadrons ($h$) in the
transverse momentum range $1.2<p_{T}<3.0$ GeV/c is measured to be
$1.19\pm0.05$(stat)$\pm0.03$(syst) in minimum-bias collisions and
shows little centrality dependence. The yield ratio of
$(p+\bar{p})/h$ in minimum-bias d+Au collisions is found to be a
factor of 2 lower than that in Au+Au collisions, indicating that
the Cronin effect alone is not enough to account for the relative
baryon enhancement observed in heavy ion collisions at RHIC.
\end{abstract}
\maketitle

%\large
{
%  The main goal of studying nuclear collisions at RHIC is to
%  understand the equation of state with partonic degree of
%  freedom~\cite{qm02proceedings}.
 Suppression of high transverse momentum ($p_{T}$) hadron production
  has been observed at RHIC in central Au+Au collisions relative to
  p+p collisions ~\cite{starhighpt,phenixhighpt,phoboshighpt,BrahmsAuAudAu}.
  This suppression has
  been interpreted as energy loss of the energetic partons traversing
  the produced hot and dense medium~\cite{jetquench}. At intermediate
  $p_{T}$, the degree of suppression depends on particle species. The
  spectra of baryons (protons and lambdas) are less suppressed than
  those of mesons (pions, kaons) ~\cite{starv2raa,phenixpid} in the
  $p_{T}$ range $2<p_{T}<5$ GeV/c. The baryon content in the hadrons
  at intermediate $p_{T}$ depends strongly on the impact parameter
  (centrality) of the Au+Au collisions with about 40\% of the hadrons
  being baryons in the minimum-bias collisions and 20\% in very
  peripheral collisions~\cite{starv2raa,phenixpid}.
  Hydrodynamics~\cite{derekhydro,pisahydro}, parton coalescence at
  hadronization~\cite{hwa,fries,ko} and gluon
  junctions~\cite{junction} have been suggested as explanations for
  the observed particle-species dependence.

  On the other hand, the hadron $p_{T}$ spectra have been observed to
  depend on the target atomic weight ($A$) and the produced particle
  species in lower energy p+A
  collisions~\cite{cronin,cronin1979,cronin1992}. This is known as the
  ``Cronin Effect'', a generic term for the experimentally observed
  broadening of the transverse momentum distributions at intermediate
  $p_{T}$ in p+A collisions as compared to those in p+p
  collisions~\cite{cronin,cronin1979,cronin1992,petersson83,accardi}. The
  effect can be characterized as a dependence of the yield on the
  target atomic weight as $A^{\alpha}$.  At energies of $\sqrt{s}
  \simeq$ 30 GeV, $\alpha$ depends on $p_{T}$ and is greater than
  unity at high $p_{T}$~\cite{cronin,cronin1979}, indicating an
  enhancement of the production cross section.  The effect has been
  interpreted as partonic scatterings at the initial
  impact~\cite{petersson83,accardi}. Thus, the Cronin effect is
  predicted to be larger in central d+Au collisions than in d+Au
  peripheral collisions~\cite{Vitev03}. At higher energies, multiple
  parton collisions are possible even in p+p
  collisions~\cite{e735kno}. This combined with the hardening of the
  spectra with increasing beam energy would reduce the Cronin
  effect~\cite{accardi}. At sufficiently high beam energy, gluon
  saturation is expected to result in a relative suppression of hadron
  yield at high $p_{T}$ in both p+A and A+A collisions and in a
  substantial decrease and finally in the disappearance of the Cronin
  effect~\cite{cgc}.

  Recent results on inclusive hadron production from d+Au collisions
  indicate that hadron suppression at intermediate $p_{T}$ in Au+Au
  collisions is due to final-state effects~\cite{BrahmsAuAudAu,stardau,otherdau}.
  The rapidity dependence of the particle yield at intermediate
  $p_{T}$ shows suppression in forward rapidity (deuteron side) and
  enhancement in the backward rapidity (Au side) in d+Au collisions at
  RHIC~\cite{asymstar,brahmsphobosdAu}.  A study of particle composition
  will help understand the origin of the rapidity
  asymmetry~\cite{hwa}.  In order to further understand the mechanisms
  responsible for the particle dependence of $p_{T}$ spectra in heavy
  ion collisions, and to separate the effects of initial and final
  partonic rescatterings, we measured the $p_{T}$ distributions of
  $\pi^{\pm}$, $K^{\pm}$, $p$ and $\bar{p}$ from 200 GeV d+Au and p+p
  collisions. In this letter, we discuss the dependence of particle
  production on $p_{T}$, collision energy, and target atomic weight.

  The detector used for these studies was the Solenoidal Tracker at
  RHIC (STAR).  The main tracking device is the Time Projection
  Chamber (TPC) which provides momentum information and particle
  identification for charged particles up to $p_{T}\sim1.1$ GeV/c by
  measuring their ionization energy loss ({\it dE/dx})~\cite{tpc}.
  Detailed descriptions of the TPC and d+Au run conditions have been
  presented in Ref.~\cite{stardau,tpc}.  A prototype time-of-flight
  detector (TOFr) based on multi-gap resistive plate chambers
  (MRPC)~\cite{startof} was installed in STAR for the d+Au and p+p
  runs. It extends particle identification up to $p_{T}\sim3$ GeV/c
  for $p$ and $\bar{p}$. In p+p and d+Au collisions, the $dE/dx$
  resolution from TPC was found to be better than $8\%$ and there is
  $2\sim3\sigma$ separation between the $dE/dx$ of pions at relativistic
  rise and the $dE/dx$ of kaons and protons at $p_{T}{}^{>}_{\sim}$2
  GeV/c~\cite{tpc}. By combining the particle identification
  capability of $dE/dx$ from TPC and Time-of-Flight from TOFr, we are
  able to extend pion identification to $\sim$3
  GeV/c~\cite{tpc,ming62}.  MRPC technology was first developed by the
  CERN ALICE group~\cite{williams} to provide a cost-effective
  solution for large-area time-of-flight coverage.

  TOFr covers $\pi/30$ in azimuth and $-1\!<\!\eta\!<\!0$ in
  pseudorapidity at a radius of $\sim220$ cm. It contains 28 MRPC
  modules which were partially instrumented during the 2003 run.
  Only particles from $-0.5\!<\!\eta\!<\!0$ are selected
  where most of the MRPC modules were instrumented.
  Each module~\cite{startof} is a stack of resistive glass plates with
  six uniform gas gaps. High voltage is applied to electrodes on the
  outer surfaces of the outer plates. A charged particle traversing
  a module generates avalanches in the gas gaps which are read out
  by 6 copper
  pickup pads with pad dimensions of $31.5\times63$
  $\mathrm{mm}^{2}$. The MRPC modules were operated at 14 kV with a
  mixture of 95\% $C_{2}H_{2}F_{4}$ and 5\% iso-butane at 1
  atmosphere. In d+Au collisions, TOFr is situated in the outgoing Au
  beam direction which is assigned negative $\eta$. The average MRPC
  TOFr timing resolution alone for the ten modules used in this
  analysis was measured to be 85 ps for both d+Au and p+p
  collisions. The ``start'' timing was provided by two identical
  pseudo-vertex position detectors (pVPD), each 5.4 m away from the
  TPC center along the beamline~\cite{pVPD}. Each pVPD consists of 3
  detector elements and covers $\sim19\%$ of the total solid angle in
  $4.4<|\eta|<4.9$~\cite{pVPD}. Due to the low multiplicity in d+Au
  and p+p collisions, the effective timing resolution of the pVPDs was
  85 ps and 140 ps, respectively.

%\pagebreak

\begin{figure}
\includegraphics*[keepaspectratio,scale=0.4]{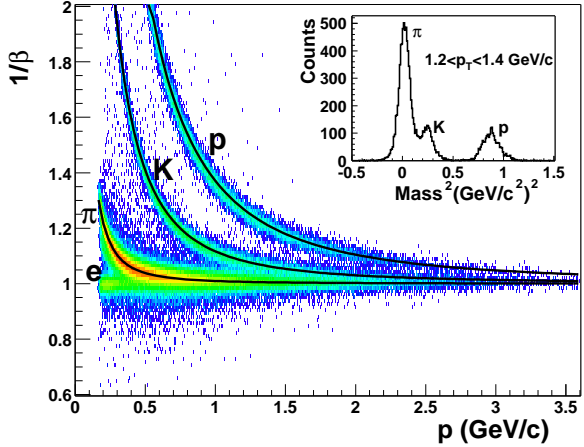}
\vspace{-0.5cm} \caption{$1/\beta$ vs. momentum for $\pi^{\pm}$,
$K^{\pm}$, and $p(\bar{p})$ from 200 GeV d+Au collisions.
Separations between pions and kaons, kaons and protons are
achieved up to $p_{T}\simeq1.6$ and $3.0$ GeV/c, respectively. The
insert shows $m^{2}=p^{2}(1/\beta^{2}-1)$ for $1.2<p_{T}<1.4$
GeV/c. Clear separation of pions, kaons and protons is seen. }
\vspace{-0.5cm} \label{Fig:beta}
\end{figure}

Since the acceptance of TOFr is small, a special trigger selected
events with a valid pVPD coincidence and at least one TOFr hit. A
total of 1.89 million and 1.08 million events were used for the
analysis from TOFr triggered d+Au and non-singly diffractive (NSD)
p+p collisions, representing an integrated luminosity of about 40
$\mathrm{{\mu}b}^{-1}$ and 30 $\mathrm{nb}^{-1}$, respectively.
The d+Au minimum-bias trigger required an equivalent energy
deposition of about 15 GeV in the Zero Degree Calorimeter in the
Au beam direction~\cite{stardau}. Minimum-bias p+p events were
triggered by the coincidence of two beam-beam counters (BBC)
covering $3.3<|\eta|<5.0$~\cite{starhighpt}. The NSD cross section
was measured to be $30.0\pm3.5$ mb by a van der Meer scan and
PYTHIA~\cite{pythia} simulation of the BBC
acceptance~\cite{starhighpt}. A small multiplicity bias (
${}^{<}_{\sim}10\%$ in d+Au and $18\%$ in p+p) at mid-rapidity was
observed in TOFr triggered events due to the further pVPD trigger
requirement and was corrected for using minimum-bias data sets and
PYTHIA~\cite{pythia} and HIJING~\cite{hijing} simulations. The
effect of the trigger bias on the mid-rapidity particle spectra
was found to be independent of particle $p_{T}$ at $p_{T}>$0.3
GeV/c~\cite{Lijuan:04}. Centrality tagging of d+Au collisions was
based on the charged particle multiplicity in $-3.8<\eta<-2.8$,
measured by the Forward Time Projection Chamber in the Au beam
direction~\cite{stardau,ftpc}. The TOFr triggered d+Au events were
divided into three centralities: most central $20\%$, $20-40\%$
and $40-\sim100\%$ of the hadronic cross section.  The average
number of binary collisions $\langle N_{bin}\rangle$ for each
centrality class and for the combined minimum-bias event sample is
derived from Glauber model calculations and listed in
Table~\ref{Tab:D}.

The TPC and TOFr are two independent systems. In the analysis,
hits from particles traversing the TPC were reconstructed as
tracks with well defined geometry, momentum, and {\it dE/dx}
~\cite{tpc}. The particle trajectory was then extended outward to
the TOFr detector plane.  Fig.~\ref{Fig:beta} shows inversed
velocity ($1/\beta$) from TOFr measurement as a function of
momentum ($p$) calculated from TPC tracking in TOFr triggered d+Au
collisions. The raw yields of $\pi^{\pm}$, $K^{\pm}$, $p$ and
$\bar{p}$ are obtained from Gaussian fits to the distributions in
$m^{2}=p^{2}(1/\beta^{2}-1)$ in each $p_{T}$ bin. For $\pi^{\pm}$
at $p_{T}>$1.8 GeV/c, an additional cut on $dE/dx$ was applied at
50\% efficiency~\cite{ming62}. The $dE/dx$ distribution was
measured by selecting on pure pion and proton samples from TOFr.
The uncertainty of this cut was evaluated by systematically
studying the yield as a function of the cut. Acceptance and
efficiency were studied by Monte Carlo simulations and by matching
TPC track and TOFr hits in real data.  TPC tracking and MRPC hit
matching efficiencies were both about $90\%$.  Weak-decay feeddown
(e.g. $K_{s}^{0}\rightarrow\pi^{+}\pi^{-}$) to pions is $\sim12\%$
at $p_{T}<$1 GeV/c and $\sim5\%$ at higher $p_{T}$, and was
corrected for using PYTHIA~\cite{pythia} and HIJING~\cite{hijing}
simulations. Inclusive $p$ and $\bar{p}$ production is presented
without hyperon feeddown correction. $p$ and $\bar{p}$ from
hyperon decays have the same detection efficiency as primary $p$
and $\bar{p}$~\cite{antiproton} and contribute about 20\% to the
inclusive $p$ and $\bar{p}$ yield, as estimated from the
simulation.

\begin{figure}
\includegraphics*[keepaspectratio,scale=0.45]{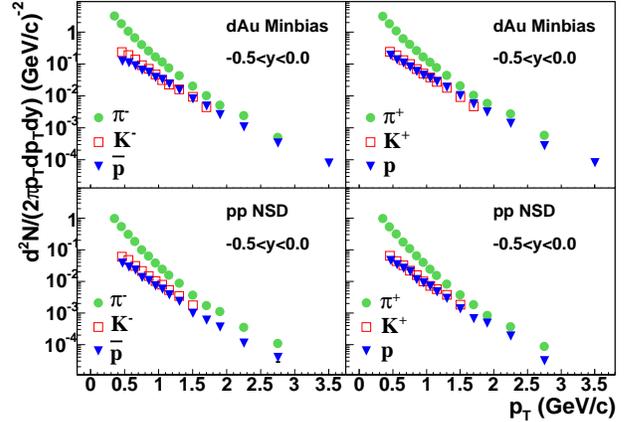}
\vspace{-0.5cm}\caption{The invariant yields of $\pi^{+}$ (filled
circles), $K^{+}$ (open squares), $p$ (filled triangles) and their
anti-particles as a function of $p_{T}$ from d+Au and NSD p+p
events at 200 GeV. The rapidity range was $-0.5<y<0.0$ with the
direction of the outgoing Au ions as negative rapidity. Errors are
statistical.} \vspace{-0.5cm}\label{Fig:spectra}
\end{figure}

 The invariant yields $d^2N/(2{\pi}p_Tdp_Tdy)$ of $\pi^{\pm}$,
$K^{\pm}$, $p$ and $\bar{p}$ from both NSD p+p and minimum-bias
d+Au events are shown in Fig.~\ref{Fig:spectra}.  The average
bin-to-bin systematic uncertainty was estimated to be of the order
of 8\%. The systematic uncertainty is dominated by the uncertainty
in the detector response in Monte Carlo simulations ($\pm7\%$).
The normalization uncertainties in d+Au minimum-bias and p+p NSD
collisions are $10\%$ and $14\%$,
respectively~\cite{starhighpt,stardau}. The charged pion yields
are consistent with $\pi^0$ yields measured by the PHENIX
collaboration in the overlapping $p_{T}$
range~\cite{phenixhighpt,otherdau}.

\begin{figure}
\includegraphics*[keepaspectratio,scale=0.45]{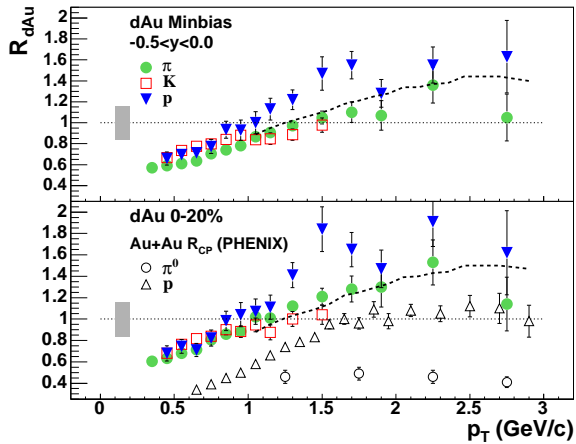}
\vspace{-0.5cm}\caption{The identified particle $R_{dAu}$ for
minimum-bias and top 20\% d+Au collisions. The filled triangles
are for $p+\bar{p}$, the filled circles are for $\pi^{+}+\pi^{-}$
and the open squares are for $K^{+}+K^{-}$. Dashed lines are
$R_{dAu}$ of inclusive charged hadrons from~\cite{stardau}. The
open triangles and open circles are $R_{CP}$ of $p+\bar{p}$ and
$\pi^{0}$ in Au+Au collisions measured by PHENIX~\cite{phenixpid}.
Errors are statistical.  The gray band represents the
normalization uncertainty of 16\%.} \vspace{-0.5cm}
\label{Fig:Rdau}
\end{figure}

Nuclear effects on hadron production in d+Au collisions are
measured through comparison to the p+p spectrum, scaled by the
number of underlying nucleon-nucleon inelastic collisons using the
ratio
\[R_{dAu}=\frac{d^{2}N/(2{\pi}p_{T}dp_{T}dy)}{T_{dAu}d^{2}\sigma^{pp}_{inel}/(2{\pi}p_{T}dp_{T}dy)}  ,\]
where $T_{dAu}={\langle N_{bin}\rangle}/\sigma^{pp}_{inel}$ describes
the nuclear geometry, and
$d^{2}\sigma^{pp}_{inel}/(2{\pi}p_{T}dp_{T}dy)$ for p+p inelastic
collisions is derived from the measured p+p NSD cross section.  The
difference between NSD and inelastic differential cross sections at
mid-rapidity, as estimated from PYTHIA~\cite{pythia}, is $5\%$ at low
$p_{T}$ and negligible at $p_{T}>1.0$ GeV/c.  Fig.~\ref{Fig:Rdau}
shows $R_{dAu}$ of $\pi^{+}+\pi^{-}$, $K^{+}+K^{-}$ and $p+\bar{p}$
for minimum-bias and central d+Au collisions. The systematic
uncertainties on $R_{dAu}$ are of the order of 16\%, dominated by the
uncertainty in normalization.  The $R_{dAu}$ of the same particle
species are similar between minimum-bias and top 20\% d+Au
collisions. In both cases, the $R_{dAu}$ of protons rise faster than
$R_{dAu}$ of pions and kaons. We observe that the spectra of
$\pi^{\pm}$, $K^{\pm}$, $p$ and $\bar{p}$ are considerably harder in
d+Au than those in p+p collisions.

The $R_{dAu}$ of the identified particles has characterstics of
the Cronin effect~\cite{cronin,cronin1979,cronin1992,accardi} in
particle production with $R_{dAu}$ less than unity at low $p_{T}$
and above unity at $p_{T}{}^{>}_{\sim} 1.0$ GeV/c. On the
contrary, the $R_{CP}$ (nuclear modification factor between
central and peripheral collisions) of identified particles in
Au+Au collisions at $\sqrt{s_{_{NN}}} = 200$ GeV as measured by
PHENIX and STAR collaborations~\cite{starv2raa,phenixpid} do not
have the above features. The $R_{CP}$ of $p+\bar{p}$ follows
binary scaling and that of $\pi^{0}$ shows large suppression of
meson production in central Au+Au collisions~\cite{phenixpid} as
depicted in the bottom panel of Fig.~\ref{Fig:Rdau}. It is notable
that the $R_{dAu}$ of proton and anti-proton are greater than
unity in both central and minimum-bias d+Au collisions while the
proton and antiproton production follows binary scaling in all
centralities in Au+Au collisions~\cite{phenixpid}.

\begin{table}{}
\caption{\label{Tab:D}$\langle N_{bin}\rangle$ from a Glauber
model calculation, $(p+\bar{p})/h$ averaged over the bins within
$1.2<p_{T}<2.0$ GeV/c (left column) and within $2.0<p_{T}<3.0$
GeV/c (right column) and the $R_{dAu}$ ratios between $p+\bar{p}$
and $h$ averaged over $1.2<p_{T}<3.0$ GeV/c for minimum-bias,
centrality selected d+Au collisions and minimum-bias p+p
collisions.  A p+p inelastic cross section of $\sigma_{inel}=42$
mb was used in the calculation. For $R_{dAu}$ ratios, only
statistical errors are shown and the systematic uncertainties are
0.03 for all centrality bins. }
{\centering
{\begin{tabular}{c|c|c|c|c} \hline \hline centrality &
 $\langle N_{bin}\rangle$ & \multicolumn{2}{c|} {$(p+\bar{p})/h$} &
 ${R_{dAu}^{p+\bar{p}}}/{R_{dAu}^h}$\\ \hline min. bias & $7.5\pm0.4$
 &$0.21\pm0.01$ &$0.24\pm0.01$ & $1.19\pm0.05$\\ 0--20\% &
 $15.0\pm1.1$ &$0.21\pm0.01$ &$0.24\pm0.02$ & $1.18\pm0.06$\\ 20--40\%
 & $10.2\pm1.0$ &$0.20\pm0.01$ &$0.24\pm0.02$ & $1.16\pm0.06$\\
 40--$\sim$100\% & $4.0^{+0.8}_{-0.3}$ &$0.20\pm0.01$ &$0.23\pm0.02$ &
 $1.13\pm0.06$\\ \hline p+p & $1.0$ &$0.17\pm0.01$ &$0.21\pm0.02$ &
 --- \\ \hline \hline
   \end{tabular}
 }
\par}
\end{table}

  Fig.~\ref{Fig:bnchratio} depicts $(p+\bar{p})/h$, the ratio of
protons $(p+\bar{p})$ over inclusive charged hadrons ($h$) as a
function of $p_{T}$ in d+Au and p+p minimum-bias collisions at
$\sqrt{s_{_{NN}}} = 200$ GeV, and Au+Au minimum-bias collisions at
$\sqrt{s_{_{NN}}} = 130$ GeV~\cite{phenixpid}. The systematic
uncertainties on these ratios were estimated to be of the order of
10\% for $p_{T}{}^{<}_{\sim}1.0$ GeV/c, decreasing to 3\% at
higher $p_{T}$. At RHIC energies, the anti-particle to particle
ratios approach unity ($\bar{p}/p=0.81\pm0.02\pm0.04$ in d+Au
minimum-bias collisions) and their nuclear modification factors
are similar. The $(p+\bar{p})/h$ ratio from minimum-bias Au+Au
collisions~\cite{phenixpid} at a similar energy is about a factor
of 2 higher than that in d+Au and p+p collisions for
$p_{T}{}^{>}_{\sim}2.0$ GeV/c.  This enhancement is most likely
due to final-state effects in Au+Au
collisions~\cite{jetquench,junction,derekhydro,pisahydro,fries,ko}.
The ratios show little centrality dependence in d+Au collisions,
as shown in Table~\ref{Tab:D}. The identified particle yields can
also provide important information and constraints for other
studies even when our measurements are in a limited rapidity range
(-0.5$<y<$0.0). Our measurement of $(p+\bar{p})/h$ ratio shows
that baryons account for only about 20\% of the total inclusive
charged hadrons with little centrality dependence. Therefore, the
measurement of rapidity asymmetry of inclusive charged hadrons
around mid-rapidity by the STAR collaboration~\cite{asymstar} is
unlikely due to a change in particle composition or baryon
stopping. For $p_{T}<2.0$ GeV/c, the $(p+\bar{p})/h$ ratio in
$\mathrm{p+\bar{p}}$ collisions at $\sqrt{s_{_{NN}}} = 1.8$
TeV~\cite{e735} is very similar to those in d+Au and p+p
collisions at $\sqrt{s_{_{NN}}} = 200$ GeV. Also shown are
$p/h^{+}$ ratios in p+p and p+W minimum-bias collisions at
$\sqrt{s_{_{NN}}} = 23.8$ GeV~\cite{cronin,cronin1979}. Although
the relative yields of particles and anti-particles are very
different, the Cronin effects are similar. At $\sqrt{s}<40$ GeV,
there is a general trend of decreasing Cronin effect of all
particles with beam energies at high
$p_{T}$~\cite{cronin1979,cronin1992}, however, the Cronin effects
of $\bar{p}$ data are less conclusive~\cite{cronin1992}.

\begin{figure}
\includegraphics*[keepaspectratio,scale=0.42]{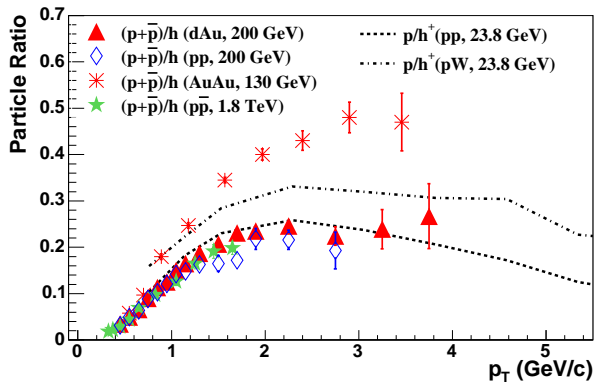}
\vspace{-0.5cm}\caption{Minimum-bias ratios of protons
($p+\bar{p}$) over inclusive charged hadrons ($h$) at
$-0.5\!<\!\eta\!<\!0.0$ from $\sqrt{s_{_{NN}}} =200$ GeV p+p (open
diamonds), d+Au (filled triangles) and $\sqrt{s_{_{NN}}} =130$ GeV
Au+Au~\cite{phenixpid} (asterisks) collisions. Results of
$\mathrm{p+\bar{p}}$ collisions at $\sqrt{s_{_{NN}}} = 1.8$
TeV~\cite{e735} are shown as solid stars. Dashed lines are results
of $p/h^{+}$ ratios from $\sqrt{s_{_{NN}}} = 23.8$ GeV p+p
(short-dashed lines) and p+W (dot-dashed)
collisions~\cite{cronin,cronin1979}. Errors are statistical. }
\vspace{-0.5cm}\label{Fig:bnchratio}
\end{figure}

The difference between $R_{dAu}$ at $\sqrt{s_{_{NN}}} = 200$ GeV
for $p+\bar{p}$ and $h$ can be obtained from the $(p+\bar{p})/h$
ratios in d+Au and p+p collisions. Table~\ref{Tab:D} shows
$R_{dAu}^{p+\bar{p}}/R_{dAu}^{h}$ determined by averaging over the
bins within $1.2<p_{T}<3.0$ GeV/c. Alternatively, we can study
Cronin effect of the identified particles by comparing the
$\alpha$ parameters of protons and pions. At lower energy, the
$\alpha$ parameter in the power law dependence on target atomic
weight $A^{\alpha}$ of identified particle production falls with
$\sqrt{s}$~\cite{cronin1979,cronin1992} at high $p_{T}$
($p_{T}\simeq4.6$ GeV/c). From the ratios of $R_{dAu}$ between
$p+\bar{p}$ and ${\pi^{+}+\pi^{-}}$, we may further derive the
$\alpha_{p+\bar{p}}-\alpha_{\pi^{+}+\pi^{-}}$ for $1.2< p_{T}<
3.0$ GeV/c to be $0.048\pm0.012$(stat)$\pm0.006$(syst). This
result is significantly smaller than the value $0.081\pm0.005$ in
the same $p_{T}$ range found at lower energies~\cite{cronin1979}.
%At lower energy, there is a general trend of decreasing $\alpha$
%with beam energies at $p_{T}\simeq$4.6 GeV/c~\cite{cronin1992}.

 In summary, we have reported the identified particle spectra of
  pions, kaons, and protons at mid-rapidity from 200 GeV p+p and d+Au
  collisions. The time-of-flight detector, based on novel multi-gap
  resistive plate chamber technology, was used for particle
  identification. The
  particle-species dependence of the Cronin effect is found to be
  significantly smaller than that from lower energy p+A collisions. In
  $\sqrt{s_{_{NN}}} = 200$ GeV d+Au collisions, the ratio of the
  nuclear modification factor $R_{dAu}$ between protons $(p+\bar{p})$
  and inclusive
  charged hadrons ($h$) in the $p_{T}$ range $1.2< p_{T}<3.0$ GeV/c
  was measured to be $1.19\pm0.05$(stat)$\pm0.03$(syst) in
  minimum-bias collisions. Both the $R_{dAu}$ values and
  $(p+\bar{p})/h$ ratios show little centrality dependence, in
  contrast to previous measurements in Au+Au collisions at
  $\sqrt{s_{NN}}$ = 130 and 200 GeV.  The ratios of protons over
  inclusive charged hadrons in d+Au and p+p collisions are found to be about a
  factor of 2 lower than that from Au+Au collisions, indicating that
  the Cronin effect alone is not enough to account for the relative
  baryon enhancement observed in heavy ion collisions.

The STAR Collaboration gratefully acknowledges the pioneering
research and development work performed by LAA project, and
especially C. Williams, under A. Zichichi on MRPC Time of Flight
Technology. We thank the RHIC Operations Group and RCF at BNL, and
the NERSC Center at LBNL for their support. This work was
supported in part by the HENP Divisions of the Office of Science
of the U.S. DOE; the U.S. NSF; the BMBF of Germany; IN2P3, RA,
RPL, and EMN of France; EPSRC of the United Kingdom; FAPESP of
Brazil; the Russian Ministry of Science and Technology; the
Ministry of Education and the NNSFC of China; Grant Agency of the
Czech Republic, FOM of the Netherlands, DAE, DST, and CSIR of the
Government of India; Swiss NSF; the Polish State Committee for
Scientific Research; and the STAA of Slovakia.


\begin{thebibliography}{9}

   \bibitem{starhighpt} STAR Collaboration, J. Adams {\it et al.}, \Journal{\PRL}{91}{172302}{2003}.

  \bibitem{phenixhighpt} PHENIX Collaboration, S.S. Adler {\it et
  al.}, \Journal{\PRL}{91}{072301}{2003}; PHENIX collaboration,
  S.S. Adler {\it et al.}, \Journal{\PRL}{91}{241803}{2003}.

  \bibitem{phoboshighpt} PHOBOS Collaboration, B.B. Back {\it et
  al.},\Journal{\PLB}{578}{297}{2004}.

  \bibitem{BrahmsAuAudAu} BRAHMS Collaboration, I. Arsene
  {\it et al.}, \Journal{\PRL}{91}{072305}{2003}.

  % STAR v2, RAA

  \bibitem{jetquench} M. Gyulassy {\it et al.},
  Review for: Quark Gluon Plasma 3, Editors: R.C. Hwa and X.N. Wang, World Scientific, Singapore, nucl-th/0302077.

  \bibitem{starv2raa} STAR Collaboration, J. Adams {\it et al.}, \Journal{\PRL}{92}{052302}{2004}.

  \bibitem{phenixpid}PHENIX Collaboration, K. Adcox {\it et al.}, \Journal{\PLB}{561}{82}{2003}; PHENIX
  Collaboration, S.S. Adler {\it et al.},
  \Journal{\PRL}{91}{172301}{2003}.

%  \bibitem{hadronquenching} K. Galmeister, C. Greiner and Z. Xu,
%  \Journal{\PRC}{67}{044905}{2003}.

  \bibitem{derekhydro} D. Teaney {\it et al.}, nucl-th/0110037;
  D. Teaney {\it et al.}, \Journal{\PRL}{86}{4783}{2001}.

  \bibitem{pisahydro} P. Huovinen, \Journal{\NPA}{715}{299c}{2003}.
  %e-Print Archive: nucl-th/0210024

\bibitem{hwa} R.C. Hwa {\it et al.}, \Journal{\PRC}{70}{024905}{2004};
 R.C. Hwa {\it et al.}, \Journal{\PRL}{93}{082302}{2004}; R.C. Hwa {\it et al.},
 \Journal{\PRC}{70}{037901}{2004}; R.C. Hwa {\it et al.},
 \Journal{\PRC}{71}{024902}{2005}.

 \bibitem{fries}R.J. Fries {\it et al.}, \Journal{\PRC}{68}{044902}{2003}. %nucl-th/0306027.

  \bibitem{ko} V. Greco {\it et al.},  \Journal{\PRL}{90}{202302}{2003}.

  \bibitem{junction}I. Vitev and M. Gyulassy, \Journal{\PRC}{65}{041902}{2002}.

%  \bibitem{qm02proceedings} Quark Matter 2002, Nucl. Phys. {\bf A715}, 1c(2003).


  % FNAL 30 GeV results

  \bibitem{cronin} J.W. Cronin {\it et al.},
  \Journal{\PRL}{31}{1426}{1973}; J.W. Cronin {\it et al.},
  \Journal{\PRD}{11}{3105}{1975}.  % model for FNAL data

  \bibitem{cronin1979} D. Antreasyan {\it et al.},
  \Journal{\PRD}{19}{764}{1979}.

  \bibitem{cronin1992} P.B. Straub {\it et al.},
  \Journal{\PRL}{68}{452}{1992}.


  \bibitem{petersson83} M. Lev and B. Petersson, Z. Phys. C {\bf 21},  155(1983).

  % models for d+A

\bibitem{accardi} A. Accardi, Contribution
to the CERN Yellow report on Hard Probes in Heavy Ion Collisions
at the LHC, hep-ph/0212148; X.N. Wang,
\Journal{\PRC}{61}{064910}{2000}.

 \bibitem{Vitev03} I. Vitev, \Journal{\PLB}{562}{36}{2003}.
  \bibitem{e735kno}T. Alexopoulos {\it et al.}, \Journal{\PLB}{435}{453}{1998}.
  % saturation model on d+A, recent
  \bibitem{cgc} D. Kharzeev {\it et al.},
  \Journal{\PLB}{561}{93}{2003}; J. Jalilian-Marian {\it et al.},
  \Journal{\PLB}{577}{54}{2003}; J.L. Albacete {\it et al.},
  \Journal{\PRL}{92}{082001}{2004}; D. Kharzeev {\it et al.},
  \Journal{\PRD}{68}{094013}{2003}; R. Baier {\it et al.},
  \Journal{\PRD}{68}{054009}{2003}.%hep-ph/0305265.

  % STAR d+Au
  \bibitem{stardau} STAR Collaboration, J. Adams {\it et al.},
  \Journal{\PRL}{91}{072304}{2003}.

  % PHENIX,PHOBOS,BRAHMS d+Au
  \bibitem{otherdau} PHENIX Collaboration, S.S. Adler {\it et al.}, \Journal{\PRL}{91}{072303}{2003}; PHOBOS
  Collaboration, B.B. Back {\it et al.},
  \Journal{\PRL}{91}{072302}{2003}.

  \bibitem{asymstar} STAR Collaboration, J. Adams {\it et al.},
  \Journal{\PRC}{70}{064907}{2004}.

  \bibitem{brahmsphobosdAu} PHOBOS Collaboration, B.B. Back {\it et al.},
  \Journal{\PRC}{70}{061901}{2004}.

  \bibitem{tpc} M. Anderson {\it et al.},
  \Journal{\NIMA}{499}{659}{2003}.

  \bibitem{startof}B. Bonner {\it et al.},
   \Journal{\NIMA}{508}{181}{2003}; M. Shao {\it et al.},
   \Journal{\NIMA}{492}{344}{2002}.

   \bibitem{ming62}STAR Collaboration, M. Shao {\it et al.}, The
   Proceedings of Hot Quarks 2004, to be published in Journal of
   Physics G, nucl-ex/0411035.


   \bibitem{williams} E. Cerron Zeballos {\it et al.},
   \Journal{\NIMA}{374}{132}{1996}; M.C.S. Williams {\it et al.},
   \Journal{\NIMA}{478}{183}{2002}.

   \bibitem{pVPD}W.J. Llope {\it et al.}, \Journal{\NIMA}{522}{252}{2004}.


   \bibitem{pythia}T. Sj\"{o}strand, P. Eden, C. Friberg {\it et al.},
   \Journal{\CPC}{135}{238}{2001}.

  \bibitem{hijing}X.N. Wang and M. Gyulassy,
   \Journal{\PRD}{44}{3501}{1991}.

    \bibitem{Lijuan:04}Lijuan Ruan, Ph.D. thesis, University of
    Science and Technology of China, 2004, nucl-ex/0503018.


    \bibitem{ftpc}K.H. Ackermann {\it et al.},
   \Journal{\NIMA}{499}{713}{2003}.

  \bibitem{antiproton}STAR Collaboration, C. Adler {\it et al.},
   \Journal{\PRL}{87}{262302}{2001}; STAR Collaboration, J. Adams {\it et al.},
  \Journal{\PRL}{92}{112301}{2004}.

   \bibitem{e735}T. Alexopoulos {\it et al.}, \Journal{\PRD}{48}{984}{1993}.

   \end{thebibliography}
 \end{document}